\documentstyle[12pt]{article}

\input{tcilatex}

\begin{document}

\title{A simplification of entanglement purification}
\author{Jin-Yuan Hsieh$^{1}$, Che-Ming Li$^{2}$, and Der-San Chuu$^{2}$ \\
$^{1}$Department of Mechanical Engineering, Ming Hsin University \\
of Science and Technology, Hsinchu 30401, Taiwan.\\
$^{2}$Institute and Department of Electrophysics, National Chiao \\
Tung University, Hsinchu 30050, Taiwan.}
\maketitle

\begin{abstract}
An idea of hybrid maps \ is proposed to establish standard entanglement
purification protocols which guarantee to purify any distillable state to a
desired maximally entangled pure state all by the standard purification
local operations and classical communications. The protocols proposed in
this work, in which two state transformations are used, perform better than
the IBM and Oxford protocols in the sense that they requir fewer operation
times in yielding a same amount of the desired pure state. One of the
proposed protocol in this work can even lead to a higher improved output
yield when it is combined with the hashing protocol, as compared with the
combined algorithm consisting of the Oxford and the hashing protocol.

\end{abstract}

Quantum information processings such as quantum teleportation\cite{1},
quantum data compression[2,3], and quantum cryptography\cite{4} rely on the
transmission of maximally entangled qubit pairs over quantum channels
between a sender(Alice) and a receiver(Bob). The quantum channel, however,
is always noisy due to the interaction with the environment and even
possibly the measurement controlled by an eavesdropper. Therefore, the pairs
shared by Alice and Bob are no longer of the desired pure ones to begin with
a quantum processing. The resource in the noisy channel then can be viewed
as a mixed state, or equivalently, an ensemble of pure states with definite
random probabilities. The probabilities of the pure states in the ensemble
are random so should be unknown to Alice and Bob in advance of a quantum
processing. Accordingly, Alice and Bob need to take an action of
entanglement purification to regain, at least asymptotically, the desired
maximally entangled pure state if the mixed state is distillable. This aim
can be achieved by Alice and Bob, using consecutive local operations and
classical communications (LOCC).

Two typical recurrence methods of entanglement purification should be
mentioned. The first entanglement purification protocol (the IBM protocol)
was developed by Bennett et. al.[5,6] in achieving a faithful quantum
teleportation. Soon later, an improved protocol entitled ''Quantum Privacy
Amplification'' (QPA, or the Oxford protocol) was addressed by Deutsch et al.%
\cite{7} in consideration of the security of a quantum cryptography over
noisy channels. Both the IBM and Oxford protocols are capable of purifying a
desired maximally entangled pure state from every distillable mixed state
whose components are not learned by Alice and Bob initially. By using the
IBM protocol, Alice and Bob can asymptotically regain the desired pure
state, but they have to consume operation time in twirling the state in
between each purification LOCC operation into a Werner state\cite{8} whose
fidelity relative to the desired pure state is always greater than $1/2$.
Compared with the IBM protocol, the Oxford protocol can provide higher
output yield, defined as the purified pairs per impure input pair,
especially when the initial fidelity with respect to the desired pure state
of the input state is close to $1/2$. In particular, the Oxford protocol is
capable of purifying any state whose average fidelity with respect to at
least one maximally entangled pure state is greater than $1/2$ and can be
directly applied to purify states which are not necessarily of the Werner
form. However, since the Oxford protocol occasionally may purify a pure
state other than the desired one, i.e., it could yield two possible pure
states, depending on the initial mixed state, Alice and Bob then are
suggested to take efforts additional to the purification LOCC operations to
transform the pure state with greatest component ($>1/2$)\ in the input
mixed state into the desired state; such action also costs operation time in
the additional local unitary operations and classical communications to
identify the mixed state and thus consumes some pairs before the standard
purification LOCC operations. The output yields induced by the IBM and
Oxford protocols are rather poor, but can be increased somewhat provided
both protocols are combined with hashing protocols, as described in refs.[5,
6]. So far, there have been modified protocols dedicated to increasing the
yield of an entanglement purification procedure, e.g., see refs. [9-11].

Surveying on these modified methods, one finds that while inducing greater
yields, they at the same time require more local unitary operations and
classical communications in the reordering schemes and hashing protocols[5,
6] that are combined in the standard purification protocols. So, when
comparing the performances of two protocols, say A and B, we can say
protocol A performs better than B either when the yield of protocol A is
greater than that of protocol B if both protocols cost equal operation
times, or when protocol A requires less operation time than protocol B
provided they induced equal yields. Instead of focusing on increasing the
yield, in this work we are intended to propose an idea of establishing
entanglement purification protocols in which the required operations are the
fewest, when compared with the standard IBM and Oxford protocols. These
protocols can purify a desired pure state by using the standard LOCC
operations alone. When using these protocols, the mixed state to be purified
needs not be transformed into the Werner state nor be reordered so that its
fidelity with respect to the desired pure state is the largest. Furthermore,
the protocols presented in this work in fact can provide better yields than
that induced by the Oxford protocol.

The standard purification LOCC operation considered in this work, as shown
in Fig. 1, should be mentioned first. In each purification LOCC operation,
Alice and Bob first perform local operations by operators $U$ and $U^{\ast }$%
, which will be defined latter, respectively. Then Alice and Bob each
performs a quantum control-not operation. They then measure the target
qubits in the computational basis, and if the outcomes, communicated via
classical channel, coincide they keep the control pair for the next step and
discard the target pair. If the outcomes do not coincide, both pairs are
discarded. In the purification LOCC operation, the state to be purified
needs not be of a Werner form. We express the mixed state in the Bell basis $%
\{\left| \Phi ^{+}\right\rangle ,$ $\left| \Psi ^{-}\right\rangle ,\left|
\Psi ^{+}\right\rangle ,$ $\left| \Phi ^{-}\right\rangle \}:$

\begin{eqnarray}
\left| \Phi ^{\pm }\right\rangle &=&\frac{1}{\sqrt{2}}(\left|
00\right\rangle \pm \left| 11\right\rangle ),  \nonumber \\
\left| \Psi ^{\pm }\right\rangle &=&\frac{1}{\sqrt{2}}(\left|
01\right\rangle \pm \left| 10\right\rangle ),
\end{eqnarray}%
where $\left| 0\right\rangle $ and $\left| 1\right\rangle $ form the
computational basis of the two-dimensional space belonging to the EPR pairs.
Let $\{a_{0},$ $b_{0},$ $c_{0},$ $d_{0}\}$ be the average initial diagonal
elements of the density operator representing the mixed state before the
protocol is begun with, and $\{a_{r},$ $b_{r},c_{r},$ $d_{r}\}$ be the
average diagonal elements of the surviving state after the $r$-th step. It
can be shown that a purification LOCC operation in fact is relative to a
nonlinear map, where the diagonal entries of the surviving state after the
LOCC operation are nonlinear functions of those before the operation.
Therefore the purification protocol considered in this work is composed of
consecutive nonlinear maps of the Bell-diagonal elements used to transform
an initial state asymptotically to a desired pure state. Suppose the state $%
\left| \Phi ^{+}\right\rangle \left\langle \Phi ^{+}\right| $ is the desired
one to be purified through the purification, we then are willing to map step
by step the initial state $\{a_{0},$ $b_{0},c_{0},$ $d_{0}\}$, where one of
the elements should be greater than $1/2$, to converge to the desired
attractor $\{1,$ $0,$ $0,$ $0\}$ as the step number $r$ is sufficiently
large. But the intrinsic property of the nonlinear map reveals that the
desired attractor is not the only one, as can be seen in the article of
Macchiavello\cite{12}, who has given the analytical convergence in the
recurrence scheme of the QPA protocol. The interesting nonlinear behavior of
the recurrence scheme in a distillation protocol is dominantly influenced by
the local unitary operations operators $U$ and $U^{\ast }$ applied by Alice
and Bob in the purification LOCC operation. Generalized expression for $U$,
controlled by two phases $\theta $ and $\phi $, is given by

\begin{equation}
U(\theta ,\phi )=\left[ 
\begin{array}{cc}
\cos (\frac{\theta }{2}) & -e^{-i\phi }\sin (\frac{\theta }{2}) \\ 
e^{i\phi }\sin (\frac{\theta }{2}) & \cos (\frac{\theta }{2})%
\end{array}%
\right] .\text{ }
\end{equation}

\FRAME{ftbpFU}{4.0447in}{1.375in}{0pt}{\Qcb{The standard purification LOCC
operations including the local controlled-NOT operation, single qubit
measurement, and local unitary operation in each party. Note that the
classical communication is not shown in this figure.}}{}{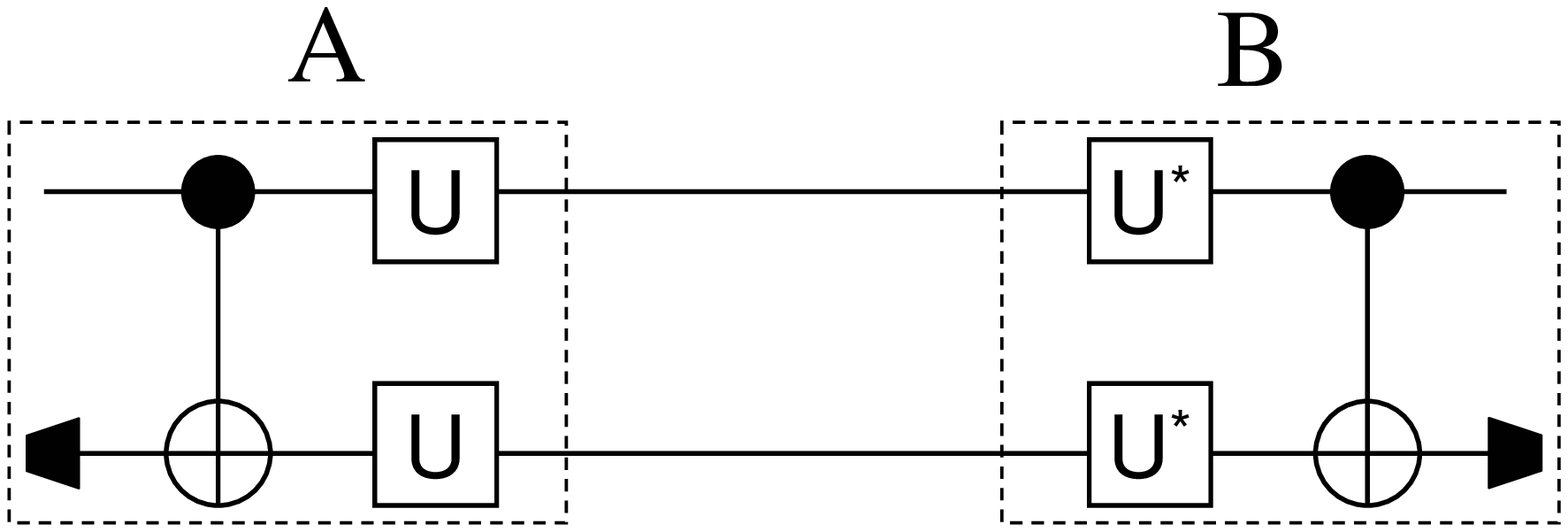}{\special%
{language "Scientific Word";type "GRAPHIC";maintain-aspect-ratio
TRUE;display "USEDEF";valid_file "F";width 4.0447in;height 1.375in;depth
0pt;original-width 7.977in;original-height 11.3887in;cropleft
"0.0377";croptop "0.7896";cropright "0.9817";cropbottom "0.5695";filename
'fig1.eps';file-properties "XNPEU";}}

It is clear that distinct choices of $\theta $ and $\phi $ will lead to
different destinations of the protocol. For example, in using the original
QPA protocol, Alice and Bob choose $\theta =\phi =\pi /2$ , i.e., they apply
the operator

\begin{equation}
U(\frac{\pi }{2},\frac{\pi }{2})=\frac{1}{\sqrt{2}}\left[ 
\begin{array}{cc}
1 & -i \\ 
i & 1%
\end{array}%
\right] .\text{ }
\end{equation}%
In this case, one will have a map $\{a_{r-1},$ $b_{r-1},c_{r-1},$ $%
d_{r-1}\}\rightarrow \{a_{r},$ $b_{r},c_{r},$ $d_{r}\}$ according to the
following relations:

\begin{eqnarray}
a_{r} &=&\frac{a_{r-1}^{2}+b_{r-1}^{2}}{p_{r-1}},b_{r}=\frac{2c_{r-1}d_{r-1}%
}{p_{r-1}},  \nonumber \\
c_{r} &=&\frac{c_{r-1}^{2}+d_{r-1}^{2}}{p_{r-1}},d_{r}=\frac{2a_{r-1}b_{r-1}%
}{p_{r-1}},\text{ for }\theta =\phi =\pi /2,
\end{eqnarray}%
where $p_{r-1}=(a_{r-1}+b_{r-1})^{2}+(c_{r-1}+d_{r-1})^{2}$ is the
probability in the $r$th step that Alice and Bob obtain coinciding outcomes
in the measurements on the target pairs ( so only $p_{r-1}/2$ of the pairs
before the $r$th step is surviving after the step ). Let us define the
domains%
\begin{eqnarray}
{\cal D}_{a} &=&\{a\in (0.5,\text{ }1];\text{ }a+b+c+d=1\}, \\
{\cal D}_{b} &=&\{b\in (0.5,\text{ }1];\text{ }a+b+c+d=1\},  \nonumber \\
{\cal D}_{c} &=&\{c\in (0.5,\text{ }1];\text{ }a+b+c+d=1\},  \nonumber \\
{\cal D}_{d} &=&\{d\in (0.5,\text{ }1];\text{ }a+b+c+d=1\},  \nonumber \\
{\cal D}_{ab} &=&{\cal D}_{a}\cup {\cal D}_{b}\text{ },  \nonumber \\
{\cal D}_{cd} &=&{\cal D}_{c}\cup {\cal D}_{d}\text{ },  \nonumber \\
\text{ \ }{\cal D}_{abcd} &=&{\cal D}_{a}\cup {\cal D}_{b}\cup {\cal D}%
_{c}\cup {\cal D}_{d}.  \nonumber
\end{eqnarray}%
In what follows we will consider the case that an initial mixed state to be
purified is in the applicable ${\cal D}_{abcd}$ because any state $\rho \in 
{\cal D}_{abcd}$ is distillable. It has been proved\cite{12} that, for the
Oxford protocol, an initial state in the domain ${\cal D}_{ab}$ will
eventually be mapped to converge to the attractor $\{1,0,0,0\}$ representing
the desired pure state $\left| \Phi ^{+}\right\rangle \left\langle \Phi
^{+}\right| $. While if the initial state is in the domain ${\cal D}_{cd}$,
then it will be mapped to approach another attractor $\{0,0,1,0\}$, or the
pure state $\left| \Psi ^{+}\right\rangle \left\langle \Psi ^{+}\right| $.
In the end, according to ref.[7], using the QPA protocol, Alice and Bob will
regain the desired pure state from any state $\rho \in {\cal D}_{abcd}$
provided they first take efforts additional to the standard purification
LOCC operations to transform the pure state $\left| \Psi ^{+}\right\rangle
\left\langle \Psi ^{+}\right| $, or $\left| \Phi ^{-}\right\rangle
\left\langle \Phi ^{-}\right| $, into the desired state $\left| \Phi
^{+}\right\rangle \left\langle \Phi ^{+}\right| $ if the input state is in
the domain ${\cal D}_{cd}$. Meanwhile, such efforts also have meaningful
implication as if the QPA is considered to be combined with the hashing
protocol[5, 6] to improve its output yield. These tedious transformations
cannot be avoided even when the input state is already in the domain ${\cal D%
}_{ab}$, because Alice and Bob initially do not have an idea about whether
the input state is exactly in the domain ${\cal D}_{ab}$ or ${\cal D}_{cd}$.
For example, if the input state has the element $c_{0}=0.7$, then Alice and
Bob should transform the state $\left| \Psi ^{+}\right\rangle \left\langle
\Psi ^{+}\right| $ into $\left| \Phi ^{+}\right\rangle \left\langle \Phi
^{+}\right| $ before the purification procedure so that the mixed state in
turn will have the element $a_{0}=0.7$.

As another example, if Alice and Bob choose $\theta =\pi /2$ and $\phi =0$,
then they have the operator

\begin{equation}
U(\pi /2,0)={\bf XH=}\frac{1}{\sqrt{2}}\left[ 
\begin{array}{cc}
1 & -1 \\ 
1 & 1%
\end{array}%
\right] ,
\end{equation}%
where ${\bf X}$ is quantum NOT gate and ${\bf H}$ is the Hadamard
transformation. Accordingly, in this case, the recurrence scheme is
described by%
\begin{eqnarray}
a_{r} &=&\frac{a_{r-1}^{2}+c_{r-1}^{2}}{p_{r-1}},b_{r}=\frac{2b_{r-1}d_{r-1}%
}{p_{r-1}},  \nonumber \\
c_{r} &=&\frac{b_{r-1}^{2}+d_{r-1}^{2}}{p_{r-1}},d_{r}=\frac{2a_{r-1}c_{r-1}%
}{p_{r-1}},\text{ for }\theta =\pi /2,\phi =0,
\end{eqnarray}%
where $p_{r-1}=(a_{r-1}+c_{r-1})^{2}+(b_{r-1}+d_{r-1})^{2}$. It should be
mentioned here that the relations (7) can also be resulted from the utility
of Hadamard transformation only, i.e., $U={\bf H}$, but this transformation
does not belong to the SU(2) operator defined in (2). Although the
analytical convergency in the recurrence scheme (7) has not yet been proved,
we find that an initial state in some domain ${\cal D}_{u}\subset {\cal D}%
_{abcd}$, which is not yet defined, will be mapped to approach the periodic
attractor representing a state interchanging step by step between $\{0.5,$ $%
0,$ $0,$ $0.5\}$ and $\{0.5,$ $0,$ $0.5,$ $0\}$, while a state in the domain 
${\cal D}_{u}^{c}$, where ${\cal D}_{u}^{c}\cup {\cal D}_{u}={\cal D}_{abcd}$%
, will be mapped to converge to the fixed attractor $\{1,$ $0,$ $0,$ $0\}$,
as wanted. For example, one can easily check to see that the initial state $%
\{0.1,$ $0.2,$ $0.6,$ $0.1\}$ will be mapped to converge to the fixed
attractor but the initial state $\{0.2,$ $0.1,$ $0.6,$ $0.1\}$, on the other
hand, will be mapped to approach the mentioned periodic attractor. So a
protocol in which the operator ${\bf XH}$ is used, unlike the QPA protocol,
will not guarantee to purify pure maximally entangled pairs.

In this work, we call a protocol a one-map protocol if Alice and Bob each
uses only one single local operator in all the purification LOCC operations,
such as the IBM and Oxford protocols. From the above examples we realize
that if only the standard purification LOCC operations are implemented, all
one-map protocols will encounter the same situation that there is always
another attractor in addition to the desired one, $\{1,$ $0,$ $0,$ $0\}$,
for a state $\rho \in {\cal D}_{abcd}$ to be mapped to converge to. This
situation thus becomes the ultimate limitation for the one-map algorithm.
Therefore, in this work, we will present a viewpoint of hybrid maps for a
purification protocol and show the fixed state $\{1,$ $0,$ $0,$ $0\}$ can be
the only attractor for an initial state $\rho \in {\cal D}_{abcd}$ to be
mapped to approach. The simple idea can be interpreted briefly. If we have
known a one-map protocol, say, controlled by $\theta _{1}$ and $\phi _{1}$,
in which a state $\rho $ belonging to some defined domain ${\cal D}%
_{1}(\subset {\cal D}_{abcd})$ can be mapped to approach the fixed attractor 
$\{1,$ $0,$ $0,$ $0\}$, then all we have to do is to find another map,
controlled by $\theta _{0}$ and $\phi _{0}$, in which a state $\rho \in 
{\cal D}_{abcd}$ will be mapped on to a subdomain of the defined ${\cal D}%
_{1}$. This kind of protocol is what we call a two-map protocol, which can
ensure Alice and Bob to regain the desired pure state $\left| \Phi
^{+}\right\rangle \left\langle \Phi ^{+}\right| $ all by using the standard
purification LOCC operations.

For the idea we have just presented, the most difficult task is the
definition of the domain ${\cal D}_{1}$. Fortunately, Macchiavello\cite{12}
has defined the domain ${\cal D}_{1}$ for the QPA protocol, in which ${\cal D%
}_{1}={\cal D}_{ab}$, as defined in (5). Therefore the QPA protocol is so
far the most convenient one-map protocol to be improved by our idea. As to
the one-map protocol described in (7), on the contrast, no definition of the
corresponding ${\cal D}_{1}$ have been proved. A concrete example of our
idea, however, will utilize these two one-map protocols. That is, in this
example the option $\theta _{1}=\pi /2$ and $\phi _{1}=\pi /2$ will be
chosen and accordingly the choice $\theta _{0}=\pi /2$ and $\phi _{0}=0$
follows. We begin with the derivation of $(1-2a_{r})$ and $(1-2c_{r})$ for $%
\theta _{0}=\pi /2$ and $\phi _{0}=0$. According to (7), we have

\begin{equation}
1-2a_{r}=\frac{(1-2a_{r-1})(1-2c_{r-1})}{p_{r-1}},\text{ }1-2c_{r}=\frac{%
(1-2b_{r-1})(1-2d_{r-1})}{p_{r-1}},
\end{equation}%
for arbitrary positive integer $r$. It is now clear to find that, since $%
p_{r-1}>0$, when $a_{0}>1/2$ or $c_{0}>1/2$, then after the first
purification LOCC operation we have $a_{1}>1/2$, while as $b_{0}>1/2$ or $%
d_{0}>1/2$, then we in turn have $c_{1}>1/2$, which consequently implies $%
a_{2}>1/2$ after the second purification LOCC operation. As a result, we
know by now that using the one-map protocol (7), we can always in two steps
map an initial state $\rho \in {\cal D}_{abcd}$ on to the domain ${\cal D}%
_{a}$, which is exactly a subdomain of ${\cal D}_{1}(={\cal D}_{ab})$ for
the standard QPA protocol. Now, we have come to the two-map protocols we
wish to present in this work. Using this two-map protocol(symbolized by
TM1), Alice and Bob have an agreement that in the first two steps of the
purification procedure, they will apply the operators $U(\pi /2,0)$ and $%
U^{\ast }(\pi /2,0)$ , respectively, to map a state $\rho \in {\cal D}%
_{abcd} $ on to the domain ${\cal D}_{a}=\{a\in (0.5,$ $1],$ $a+b+c+d=1\}$,
and then they will apply the standard QPA operators $U(\pi /2,\pi /2)$ and $%
U^{\ast }(\pi /2,\pi /2)$ to purify the surviving state to the desired state 
$\left| \Phi ^{+}\right\rangle \left\langle \Phi ^{+}\right| $ in the rest
purification LOCC operations. Interestingly, an alternative two-map protocol
(symbolized by TM2) can be used as well, in which the operators $U(\pi /2,0)$
and $U^{\ast }(\pi /2,0)$ are applied only at the second purification LOCC
operation, since after the first LOCC operation, in which the QPA operators $%
U(\pi /2,\pi /2)$ and $U^{\ast }(\pi /2,\pi /2)$ are used, the state has
been mapped on to the domain ${\cal D}_{ac}$\cite{12}.\FRAME{fhFU}{2.9776in}{%
3.5665in}{0pt}{\Qcb{The variations of the yield and the comparing purity (in
the inserted diagram) at ten times of the recurrence method.}}{}{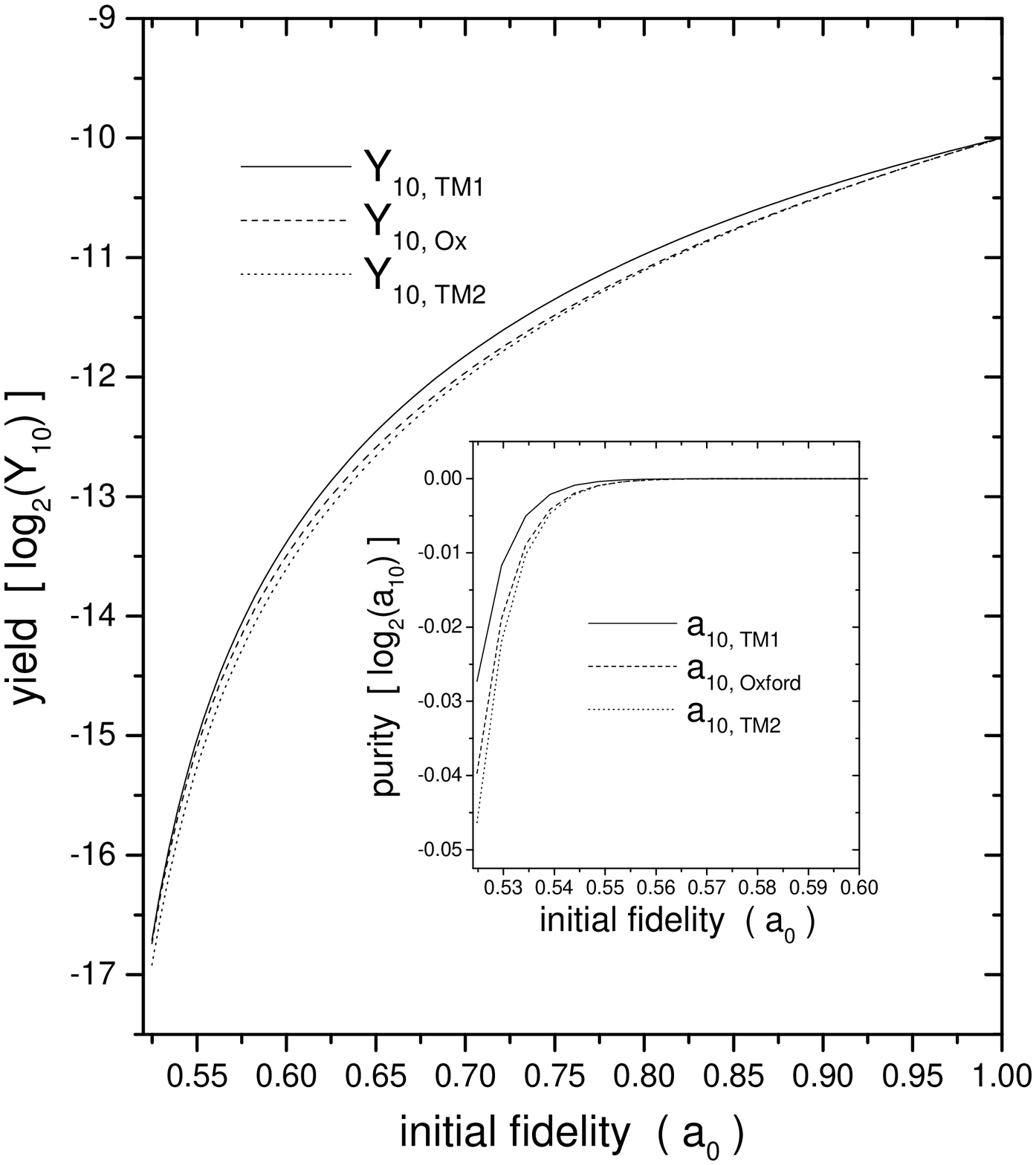}{%
\special{language "Scientific Word";type "GRAPHIC";maintain-aspect-ratio
TRUE;display "USEDEF";valid_file "F";width 2.9776in;height 3.5665in;depth
0pt;original-width 7.977in;original-height 11.3887in;cropleft
"0.0042";croptop "0.8463098";cropright "0.9416";cropbottom
"0.0585098";filename 'fig2.eps';file-properties "XNPEU";}}\FRAME{fhFU}{%
2.917in}{3.6461in}{0pt}{\Qcb{The variations of the yield and the comparing
purity (in the inserted diagram) at five times of the recurrence method.}}{}{%
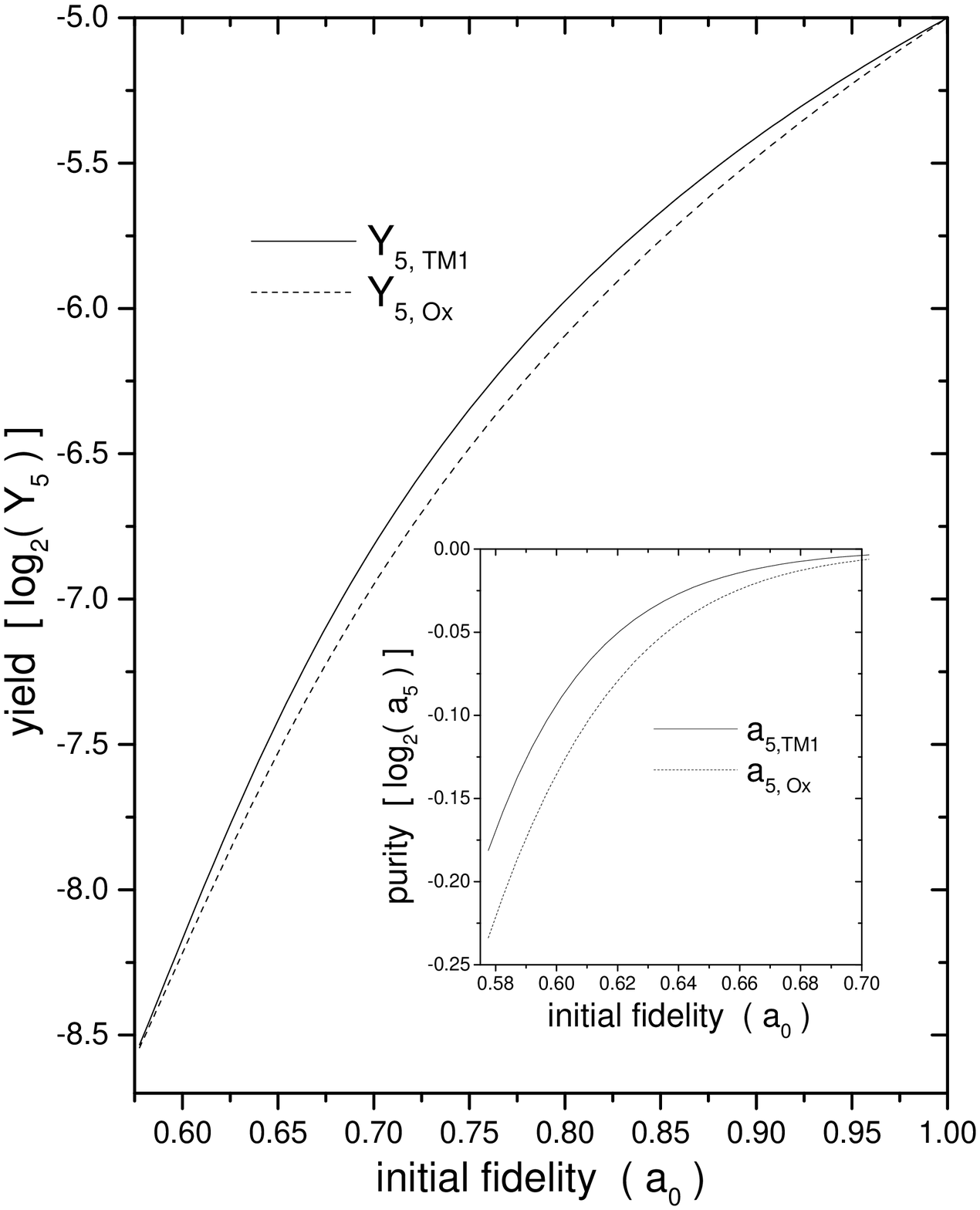}{\special{language "Scientific Word";type
"GRAPHIC";maintain-aspect-ratio TRUE;display "USEDEF";valid_file "F";width
2.917in;height 3.6461in;depth 0pt;original-width 7.977in;original-height
11.3887in;cropleft "0.0152057";croptop "0.8769489";cropright
"0.9398057";cropbottom "0.0653489";filename 'fig3.eps';file-properties
"XNPEU";}}

Apparently, our protocols TM1 and TM2 are composed of only the standard
purification LOCC operations, without using any additional local operations
and classical communications in transforming the mixed state into a Werner
state, as needed in the IBM protocol, or transforming one of the Bell states
whose fidelity is the largest into the desired pure state $\left| \Phi
^{+}\right\rangle $ in advance of the Oxford operations. Therefore the
fewest operations are required in our purification algorithms, as compared
with the IBM and Oxford protocols. Furthermore, when comparing the output
yields and the fidelities (or purities) produced by the IBM, the Oxford, and
our two-map protocols, we find the protocol TM1 can provide better yields
and fidelities than the Oxford protocol (which performs better than the IBM
protocol), while the protocol TM2 can perform almost equally to the Oxford
protocol, although this is not the primary purpose of this work. In our
numerical simulations, the yield, or the fraction of the surviving pairs,
defined by $Y_{r}=p_{0}p_{1}...p_{r-1}(2^{-r}),$ where $r$ denotes the
iteration number, was first computed up to $r=10$ for each input state to be
purified. The variations of the yield as functions of the initial fidelity $%
a_{0}$ are shown in Fig. 2, in which (and also in the following figures)
each yield (and each purity) was the average value computed over ten
thousand random states possessing the same initial fidelity. The
corresponding purities after the ten iterations are also shown in Fig. 2. It
is shown that although, after the ten iterations, the resulted purities
produced by using the Oxford, TM1, and TM2 are high, the yields of them are
rather poor, especially when the initial fidelity is close to $1/2$. The
yield, however, can be further improved by combining the recurrence method
with the hashing protocol[5, 6] as long as the purity is high enough (e.g.
higher than $0.8107$ for a Werner state) when the recurrence scheme is
performed in only a few iterations. In Fig. 3 we show the yields $Y_{5}$ and
the correspond purities $a_{5}$ produced by the Oxford and the TM1 protocols
after five iterations, respectively. This figure shows that when the initial
fidelities are greater than some specific values near $1/2$ for both cases
(of course the specific fidelities can be lowered if the iteration is
increased), the hashing protocols then are applicable after the five
iterations in running the recurrence schemes. Fig.3 shows that after the
five iterations, the surviving fraction $Y_{5,TM1}$ and the corresponding
purity $a_{5,TM1},$ produced by the the TM1 protocol are slightly higher
than the surviving fraction $Y_{5,Ox}$ and the purity $a_{5,Ox}$, which are
resulted from using the Oxford protocol. The slight differences in $Y_{5}$
and $a_{5}$, however, can induce significant difference between the improved
yields when the hashing protocol is switched on after the five iterations.
The evidence can be seen in Fig. 4, in which both the improved yields $%
Y_{5,TM1}^{\prime }$ and $Y_{5,Ox}^{\prime }$ and the ratio of the improved
yields ($Y_{5,TM1}^{\prime }/Y_{5,Ox}^{\prime }$) as functions of the
initial fidelity are shown; the improved yield is defined by $Y_{r}^{\prime
}=Y_{r}(1-S(\rho _{r}))$, where $\rho _{r}$ is the von Neumann entropy of
the surviving mixed state $\rho _{r}$. It is clearly shown in Fig. 4 that
the ratio $Y_{5,TM1}^{\prime }/Y_{5,Ox}^{\prime }$ becomes greater as the
initial fidelity is closer to $1/2$.\FRAME{fhFU}{2.9732in}{3.7689in}{0pt}{%
\Qcb{The variations of the improved yields $Y_{5,TM1}^{\prime }$ and $%
Y_{5,Ox}^{\prime }$ and the comparing ratio $(Y_{5,TM1}^{\prime
}/Y_{5,Ox}^{\prime })$ (in the inserted diagram).}}{}{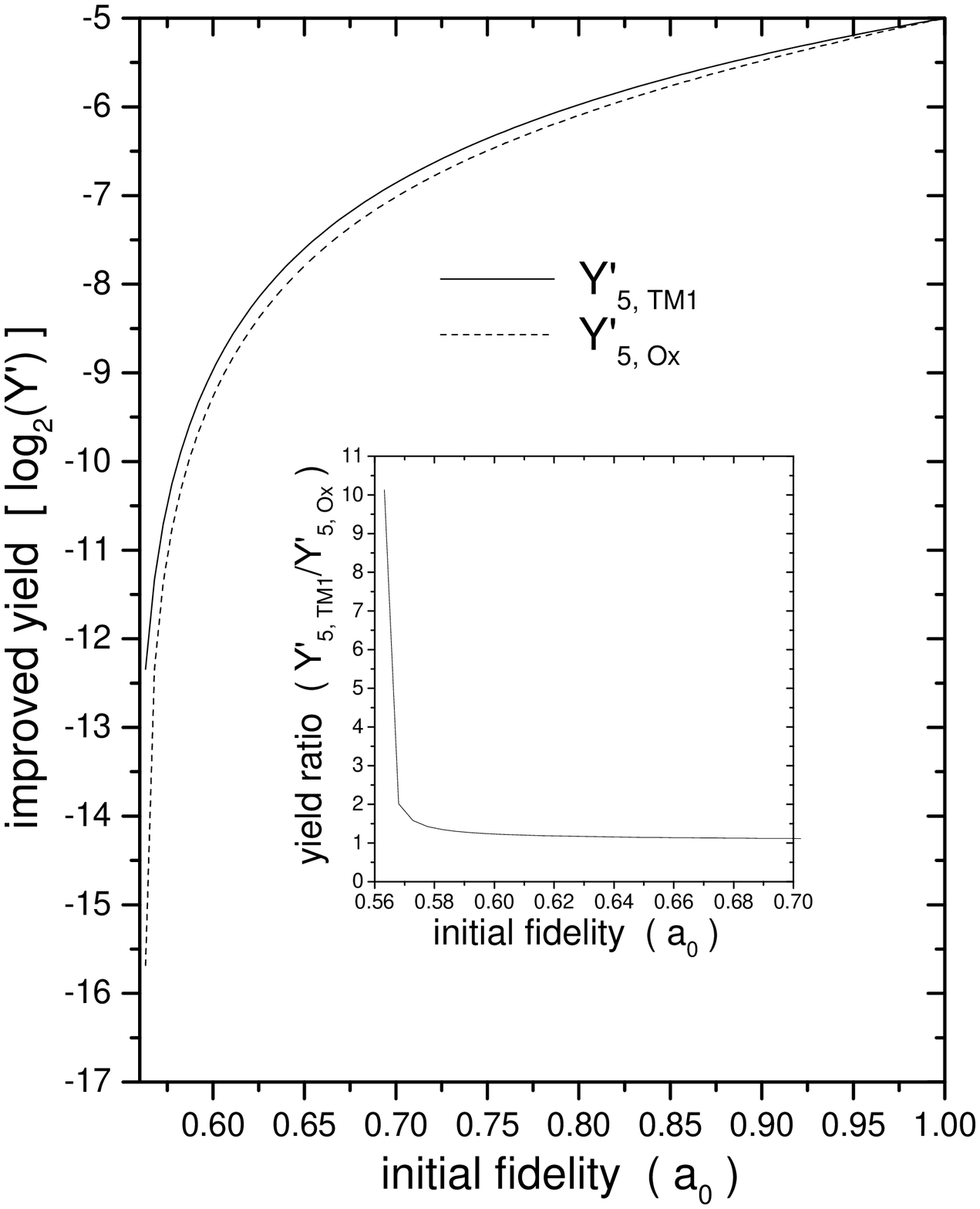}{\special%
{language "Scientific Word";type "GRAPHIC";maintain-aspect-ratio
TRUE;display "USEDEF";valid_file "F";width 2.9732in;height 3.7689in;depth
0pt;original-width 7.977in;original-height 11.3887in;cropleft
"0.0397441";croptop "0.8560716";cropright "0.9176441";cropbottom
"0.0742716";filename 'fig4.eps';file-properties "XNPEU";}}

To summarize, in the recurrence scheme of a one-map entanglement
purification protocol, the nonlinear behavior of the four Bell-diagonal
elements of the density matrix representing the mixed state to be purified
reveals that there is always another attractor other than the desired fixed
attractor. This indicates that not all the distillable input state can be
purified to the desired maximally entangled pure state all by the standard
purification LOCC operations in a one-map protocol. Therefore some tedious
efforts additional to the purification LOCC operations are needed in using
the typical IBM and Oxford protocols to purify a desired pure state from any
distillable state. The proposed two-map purification protocols TM1 and TM2,
on the contrast, can guarantee that all the distillable input states can be
purified to the desired pure state all by the standard purification LOCC
operations. That an entanglement purification can be accomplished all by the
standard purification LOCC operations is crucially important to a
significant improvement for the purification process. By such improvement,
we then do not have to identify the mixed state and consequently do not
consume any pairs before the purification LOCC operations. The proposed
two-map protocols perform better than the one-map IBM and Oxford protocols
in the sense that they require the least operation times in yielding a same
amount of useful EPR pairs. Surprisingly, the protocol TM1 is found able to
induce higher yields and purities than the Oxford protocol. This is
crucially important as the hashing protocol is combined with the recurrence
algorithm to improve the output yield. The proposed two-map protocols,
however, like the standard IBM and Oxford protocols, should be implemented
if the initial state possesses a fidelity very close to $1/2$ only after
enhancing the state's fidelity. For example, it has been shown\cite{13} that
only inseparable two-qubit state with ''free'' entanglement, however small,
can be distillable to a pure form by using local filtering[14, 15] to
enhance the state's fidelity first. An interaction with the environment\cite%
{16} can even be allowed to enhance the fidelity of a quantum teleportation.
The fidelity enhancement, however, is not an issue to be concerned with in
this work.

\end{document}